\documentclass[]{aastex631}

\shorttitle{Climate Regimes Across the Habitable Zone}
\shortauthors{Lobo et al.}
\graphicspath{{./}{figures/}}

\usepackage[colorinlistoftodos]{todonotes} 
\usepackage{gensymb}
\usepackage{soul}

\begin{document}

\title{
Climate Regimes Across the Habitable Zone: \\
a Comparison of Synchronous Rocky M- and K-dwarf Planets
} 

\author[0000-0003-3862-1817]{Ana H. Lobo}
\affiliation{University of California, Irvine \\
Department of Physics \& Astronomy \\
4129 Frederick Reines Hall \\
Irvine, CA 92697-4575, USA}

\author[0000-0002-7086-9516]{Aomawa L. Shields}
\affiliation{University of California, Irvine \\
Department of Physics \& Astronomy \\
4129 Frederick Reines Hall \\
Irvine, CA 92697-4575, USA}
\affiliation{NASA NExSS Virtual Planetary Laboratory, Seattle, WA}



\begin{abstract}

M- and K-dwarf stars make up 86\% of the stellar population and host many promising astronomical targets for detecting habitable climates in the near future. Of the two, M dwarfs currently offer greater observational advantages and are home to many of the most exciting observational discoveries in the last decade. But K dwarfs could offer even better prospects for detecting habitability by combining the advantage of a relatively dim stellar flux with a more stable stellar environment. Here we explore the climate regimes that are possible on Earth-like synchronous planets in M- and K-dwarf systems, and how they vary across the habitable zone. We focus on surface temperature patterns, water availability, and implications for habitability. We find that the risk of nightside cold-trapping decreases with increased orbital radius and is overall lower for K-dwarf planets. With reduced atmospheric shortwave absorption, K-dwarf planets have higher dayside precipitation rates and less day-to-night moisture transport, resulting in lower nightside snow rates. These results imply a higher likelihood of detecting a planet with a moist dayside climate in a habitable ``eyeball” climate regime orbiting a K-dwarf star. We also show that ``terminator habitability” can occur for both M- and K-dwarf land planets, but would likely be more prevalent in M-dwarf systems. Planets in a terminator habitability regime tend to have slightly lower fractional habitability, but offer alternative advantages including instellation rates more comparable to Earth in regions that have temperatures amenable to life.

\end{abstract}

\keywords{planets and satellites: terrestrial planets --- planets and satellites: atmospheres}


\section{Introduction} \label{sec:intro}

Dim stars are favored in the search for life beyond the Solar System because they allow for larger planet-star contrasts and enhanced signal-to-noise ratios in observations, such that M dwarfs have been frequently prioritized in both observational and modeling research. Rocky M-dwarf planets are among some of the most exciting observational targets currently available, as has been emphasized by recent observations of the TRAPPIST-1 system \citep[e.g.][]{Lincowski2023}. They are also exceptionally abundant targets, given that M dwarfs make up $\sim$70\% of all stars \citep{Bochanski2010} and are expected to have abundant rocky Earth-sized planets \citep{Mulders2015}. 

However, M-dwarf stellar properties pose many challenges for developing and sustaining habitable conditions, such that their prospects for habitability are debated \citep{Scalo2007, Tarter2007, Shields2016}. Water is a minimum requirement for life as we know it, such that accumulating and retaining water is often regarded as a minimum condition for planetary habitability \citep{Kasting1993, Kopparapu2013}. However, water retention on M-dwarf planets may be especially difficult. M dwarfs are extremely active stars \citep{Reid2005, Scalo2007}, with intense X-ray and UV radiation output during the first billion years of their evolution, and persistent flare activity throughout their lifetimes \citep{Segura2010}. Enhanced heating during the M dwarf's protracted protostar phase could promote planet-wide desertification \citep{Bolmont2017, Luger2015}. Even if the planet remains or becomes water rich during later stages, M-dwarf planets can also be subject to rapid nightside cold-trapping and water vapor escape, such that many planets in the habitable zone could have arid dayside climates \citep{Lobo2023}. Therefore, it is important that we explore the possibly of more water-limited planetary scenarios, which may be especially common among M-dwarf planets, and assess their prospects for habitability. 

In contrast, the K-dwarf pre-main sequence phase is an order of magnitude shorter \citep{Luger2015}, with young K-dwarf planets receiving significantly smaller UV and X-ray fluxes than those orbiting early M dwarfs \citep{Richey2019}. The reduced high-energy radiation would help preserve the planetary atmosphere and water reservoirs. K-dwarf planets are also expected to receive enhanced volatile delivery \citep{Ciesla2015}, such that K-dwarf habitable zone planets could more easily obtain and retain large water reservoirs during their earlier evolution. 

K dwarfs combine the observational advantages of a dim star with a more stable long-term stellar flux, making them excellent targets for habitability studies. Similarly to M dwarfs, they also offer advantages over brighter stars like the Sun. K dwarfs have longer lifespans, allowing extra time for life to develop. Biosignatures on K-dwarf planets may also be easier to detect. For example, less observing time would be needed to detect NO$_2$ \citep{Kopparapu2021NO2}. Also, the photochemical lifetime of methane in the presence of oxygen is enhanced on K-dwarf planets and they are expected to have stronger oxygen and methane features in reflected-light spectra than equivalent G-dwarf planets, allowing for shorter observing times \citep{Arney2019}. K-dwarf stars are also relatively abundant, representing $\sim$13\% of the Milky-way's population, roughly twice the number of Sun-like stars. Though still uncertain, K-dwarf planet occurrence rates for small planets (1-4 $R_\oplus$) in the inner disk are expected to be relatively high, surpassing local G-dwarf rates \citep{Mulders2015planetfreq}.

We expect an increase in K-dwarf planet detections with the James Webb Space Telescope \citep{ Gardner2006, Kalirai2018}, and these planets will become increasingly easier to observe with the next generation of space-based telescopes, which might resemble LUVOIR \citep{luvoirfinalreport} and HabEx \citep{habexfinalreport}. These future missions offer exciting prospects for K-dwarf biosignature detection, and it is expected that in just a 2-year search, a LUVOIR-like telescope could detect 8-12 habitable planet candidates orbiting K-dwarf stars \citep{luvoirfinalreport}. 

Despite being considered exceptionally promising hosts for habitable planets \citep{Cuntz2016, Arney2019}, the climate of K-dwarf planets have received substantially less attention than their M-dwarf counterparts. Here, we build on the analysis from \citet{Lobo2023} to explore the climate of synchronously rotating M- and K-dwarf planets across the habitable zone, with special focus on K-dwarf planets. We compare two potentially habitable climate regimes. The first is an ``eyeball" climate regime, which is characterized by moderate temperatures on the dayside, with the habitable region centered at the substellar point, and freezing temperatures on the nightside. The second is the ``terminator habitability" climate regime, where a temperate habitable region exists near the terminator of a synchronously rotating planet, while dayside temperatures exceed habitable limits and nightside temperatures are below freezing. 

In this paper, we focus on characterizing planetary dayside climates and water availability, while also exploring possible advantages for habitability on K-dwarf planets. We begin with an overview of topics most relevant to the inner habitable zone (section~\ref{sec:InnerHabZone}), including cloud coverage (\ref{sub:Clouds}) and changing day-night temperature contrasts (\ref{sub:delT}). We then explore the possibility of a terminator habitability climate regime on K-dwarf planets (\ref{TH}) and how the water cycle and long term prospects for water availability on M- and K-dwarf planets vary across the habitable zone (\ref{WaterCycle}). Section \ref{Discussion} provides a broader discussion of climate regimes associated with the inner and outer habitable zones, and section \ref{sec:Conclusions} summarizes our conclusions.

\section{Methods} \label{sec:Methods}

Simulations were run with ExoCAM \citep{githubExoCAM, Wolf2022}, which is a modified version of the Community Atmosphere Model (CAM4), with correlated-k radiative transfer (ExoRT), and a finite-volume dynamical core. A complete description of the code and its lineage is available in \citet{Wolf2022} and references within. The simulations have Earth-like planetary properties including radius and gravity, with an atmospheric composition of 40 Pa CO$_2$, 0.17 Pa CH$_4$, and the remainder N$_2$, totaling 1 bar for the dry component of the atmosphere, with variable H$_2$O also included. Planets in the habitable zones of dimmer stars are expected to have their obliquities substantially reduced by tidal forces \citep{Heller2011}, such that we can simplify our experiment setup and assume obliquity is zero. We also assume eccentricities will be small, and set them to zero. 

We use M-dwarf planet simulations from \citet{Lobo2023}, with the host star being AD Leonis (M3.5V). We also perform ExoCAM simulations for K-dwarf planets with the same planetary properties orbiting the host star Epsilon Eridani (HD22049, K2V). We note that AD Leonis is a relatively bright M-dwarf star, such that synchronously rotating planets in its habitable zone are in a slowly rotating regime, as is the case for the synchronous HD22049 planets that are even further from their host star. The relationship between the orbital radius and orbital period (which is equal to the planetary rotation rate) is set by Kepler's third law. 

To compare M- and K-dwarf planets, we provide simulations with similar solar constants, such that the amount of energy reaching a given M- or K-dwarf planet is similar. However, the radiation follows the spectral profile of the host star, such that M-dwarf planets receive a greater fraction of their energy at near-IR wavelengths. We also prescribe the two-band snow and ice albedos, which we refer to as the visible ($0.25 < \lambda < 0.75 \mu m$) and near-IR ($0.75 < \lambda < 2.5  \mu m$), weighted by the stellar spectrum (Table.~\ref{tab:albedovalues}.) We note that the exact values of the solar constants are not identical between M- and K-dwarf planets because the orbital period length in days must be an integer in the model. Simulations are named as a function of their host star type, their surface type, and their solar constant (Table \ref{tab:data_list}). For example, an M-dwarf simulation (M) in land planet configuration (ld) receiving 1629 W/m$^2$ will be labeled Mld1600. We round the instellation values to make the notation easier to read.

\begin{deluxetable}{c c c}[h]
\tablenum{1}
\tablecaption{Albedo Values \label{tab:albedovalues}}
\tablewidth{0pt}
\tablehead{
\colhead{ } & \colhead{Snow} & \colhead{Ice}
  }
\startdata
M-dwarf (Visible/Near-IR) & 0.97/0.48  & 0.64/0.17 \\ 
K-dwarf (Visible/Near-IR) & 0.98/0.57  & 0.70/0.21 \\
\enddata
\end{deluxetable}
\vspace{-13mm}

All simulations have a horizontal resolution of $4\degree \times 5\degree$, 40 vertical levels, and a time step of 1800s, and are run until they reach steady state. Results are shown as a time average of the last 10 years. The only exception is Maq2100, which does not reach steady state due to numeric instabilities as it entered a runaway greenhouse state. Values pertaining to this simulation were taken using the average of the last stable month of model time. 

Results are primarily shown on a planetary grid where the substellar point is at 0\degree \space longitude and 0\degree \space latitude (e.g. Fig.~\ref{fig:surfaceTemperature}). Figures with a vertical profile use the CESM hybrid sigma-pressure coordinate, plotted on a scale of 0-1 where 1 is the simulation's reference surface pressure (P$_o$). For plots showing stream functions, these were calculated by first switching to a planetary grid where the substellar point was at the northern pole (as defined in \citealt{Koll2015}). We will refer to this projection as the ``Tidally locked" projection. With this projection, we are able to take a zonal integral that forms a circle centered on the substellar point and better capture the overturning circulation. We use the same stream function definition as \citep{Hammond2021}, where: 

\begin{equation}
    \Psi' = \frac{2 \pi \cos(\theta')}{g} \int_{0}^{p} [v'] dp
\end{equation}
where $\Psi'$ is the meridional stream function, $a$ is the planetary radius, $\theta'$ is the angle towards the antistellar point, $p$ is pressure, and $[ \; ]$ represents an average over longitude in the new coordinates (see details in \citealt{Hammond2021}).

\begin{deluxetable}{cccllc}[h]
\tablenum{2}
\tablecaption{List of Simulations \label{tab:data_list}}
\tablewidth{0pt}
\tablehead{
\colhead{Name} & \colhead{Host} & \colhead{Type} & \colhead{$P_{\text{orb}}$ (days)} & \colhead{a (au)} &
\colhead{F (W/m$^2$) }
}
\startdata
\multicolumn{6}{c}{M-dwarf planets} \\
Maq1100 & M & Aquaplanet & 41 days & 0.174 & 1073 \\
Maq1400 & M & Aquaplanet & 34 days & 0.154 & 1378 \\ 
Maq1600 & M & Aquaplanet & 30 days & 0.141 & 1629 \\ 
Maq2100 & M & Aquaplanet & 25 days & 0.125 & 2077 \\  
Mld1400 & M & Q=0.0001 & 34 days & 0.154 & 1378 \\ 
Mld2100 & M & Q=0.0001 & 25 days & 0.125 & 2077 \\ 
\multicolumn{6}{c}{K-dwarf planets} \\
Kaq1100 & K & Aquaplanet & 212 days & 0.652 & 1089 \\
Kaq1400 & K & Aquaplanet & 180 days & 0.584 & 1355 \\
Kaq1600 & K & Aquaplanet & 156 days & 0.531 & 1639 \\
Kaq2000 & K & Aquaplanet & 132 days & 0.475 & 2048 \\
Kaq2700 & K & Aquaplanet & 107 days & 0.413 & 2710 \\
Kld1400 & K & Q=0.0001 & 180 days & 0.584 & 1355 \\
Kld2000 & K & Q=0.0001 & 132 days & 0.475 & 2048 \\
\enddata
\end{deluxetable}

To facilitate comparisons, it is convenient to define the habitable zone limits as a function of incident stellar flux rather than orbital distance. The inner habitable zone limit has two commonly used definitions: I. the moist greenhouse limit (MGL) and II. the runaway greenhouse limit. The former is the limit most relevant for habitability, given that planets inward of the MGL would lose their water to atmospheric escape. Based on the limits calculated by \citet{Kopparapu2013}, the inner habitable zone limit for our M- and K-dwarf planets would be 0.9$S_\odot$ and 1$S_\odot$ respectively. However, their model did not include the effects of clouds and, as expected, we find that the MGL is reached at higher instellation values for our GCM simulations.

The outer habitable zone limit is commonly defined based on the CO$_2$ ``maximum greenhouse limit". In \citet{Kopparapu2013}, it is calculated based on the minimum stellar flux needed to maintain a global mean surface temperature of 273 K, while varying CO$_2$ to obtain the maximum greenhouse effect. Based on their results, we would expect the outer habitable zone limit to be 0.2$S_\odot$ and 0.3$S_\odot$, for M- and K-dwarfs respectively. However, this definition is not ideal for us given that we are interested in the dayside temperature range and not the mean planetary climate. It is also not directly applicable to our simulations, given that we are keeping CO$_2$ values constant. For our purposes, the outer habitable zone limit is defined by dayside fractional habitability reaching zero.

The exact position of the inner and outer habitable zone limits are dependent on planetary properties \citep[e.g.][]{Yang2019radius, Thomson2019, Meadows2018, Kasting1993, Yang2013, Yang2014HZRotation}, including atmospheric CO$_2$ \citep{Kopparapu2013}. We do not seek to pinpoint the position of these limits. For our simulations, we use Earth-like planetary properties, with the atmospheric composition prescribed to pre-industrial values. We also do not include an active carbonate-silicate cycle \citep{Walker1981}, which could significantly impact the climate on these planets as silicate weathering rates, and subsequently atmospheric CO$_2$ concentrations, would vary across the habitable zone. It is also important to note that the typical inner habitable zone limits are calculated assuming abundant planetary water. For water-limited planets, the innerlimit occurs at significantly higher instellation \citep{Abe2011}, and we include a comparison of aquaplanet and land planet simulations to explore the impact of water abundance on planetary climate.

The aquaplanet simulations (``aq") have uniform slab oceans with 50m mixed layer depth and no ocean energy transport. Land planet (``ld") simulations have a sandy surface everywhere and were initialized with a limited amount of water homogeneously distributed in the atmosphere, by prescribing the specific humidity Q = 0.0001 kg/kg. The simulation configurations were selected to match those of \citet{Lobo2023} and all simulations can be found on Zenodo \citep{LoboData2024}.

\section*{Results} \label{sec:Results}

\section{The Inner Habitable Zone}\label{sec:InnerHabZone}

Previous studies have shown that the inner edge of the habitable zones of brighter stars have a higher stellar flux limit due, in large part, to the relationship between the stellar spectral energy distribution and planetary wavelength-dependent albedos \citep{Rushby2020}, as well as wavelength-dependent molecular absorption properties \citep[e.g.][]{Kopparapu2013}, which lead to significant changes in key radiative climate feedbacks \citep{Shields2013, Shields2014, Shields2019, Kopparapu2016, Wolf2017, Turbet2016, Komacek2019}. As expected, this behavior is evident in our simulations. For example, while an M-dwarf Earth-like aquaplanet receiving approximately 20\% more radiation than Earth is in a moist greenhouse state, and a simulation receiving 50\% more radiation is in a runaway greenhouse state, K-dwarf simulations receiving the same amount of stellar radiation do not reach the MGL. 

K-dwarf aquaplanets have higher dayside albedos, due both to a higher reflectivity of clouds (Fig.~\ref{fig:Clouds}), and higher reflectivity of the sea ice (Table \ref{tab:albedovalues}). For the same stellar constant there is less energy entering the climate system. This can be confirmed by comparing the top-of-atmosphere (TOA) reflected stellar radiation (Fig.~\ref{fig:NEI_M} and \ref{fig:NEI_K}, orange lines) and remaining downward stellar radiation (yellow lines). The differences are especially prominent for warmer planets (e.g. aq1400, aq1600, and aq2000). The details of the cloud behavior are addressed in the following subsection (\ref{sub:Clouds}). 

\begin{figure}[h]
\centering
\includegraphics[width=0.9\textwidth]{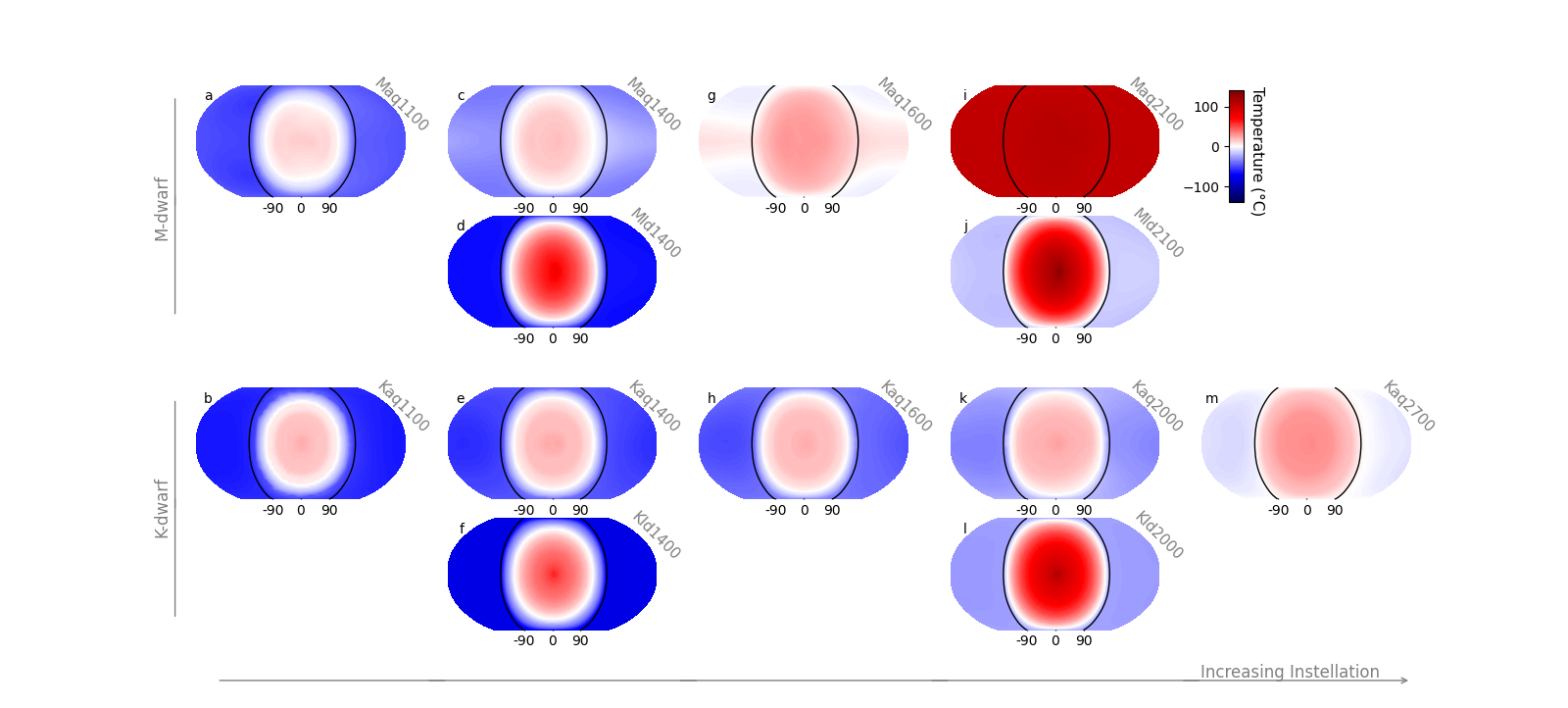}
\caption{\label{fig:surfaceTemperature} Planetary surface temperatures (\textdegree C) on M- and K-dwarf planets. Substellar point is plotted at the center, and black lines indicate the terminator. The top rows show aquaplanets (aq) and bottom rows show land planets (ld).}
\end{figure}

All of our K-dwarf aquaplanet simulations are in what is casually referred to as an ``eyeball" climate regime. An ``eyeball" climate is characterized by moderate temperatures on the dayside, with the habitable region centered at the substellar point, and freezing temperatures on the nightside. For these Earth-like aquaplanets, the nightide remains freezing in all simulations that do not reach the MGL, as can be noted in Fig.~\ref{fig:surfaceTemperature}. Among the M-dwarf simulations, Maq1600 has surpassed the MGL and only a portion of the nightside ocean is covered in sea ice, while Maq2100 is in a runaway greenhouse state and becomes numerically unstable. For the K-dwarf planets, Kaq2700 is the only one that reaches the MGL, but even so the nightside temperatures are still below freezing. 

\subsection{Clouds on K- vs M-dwarf planets}\label{sub:Clouds}

\begin{figure}[h]
\centering
\includegraphics[width=0.9\textwidth]{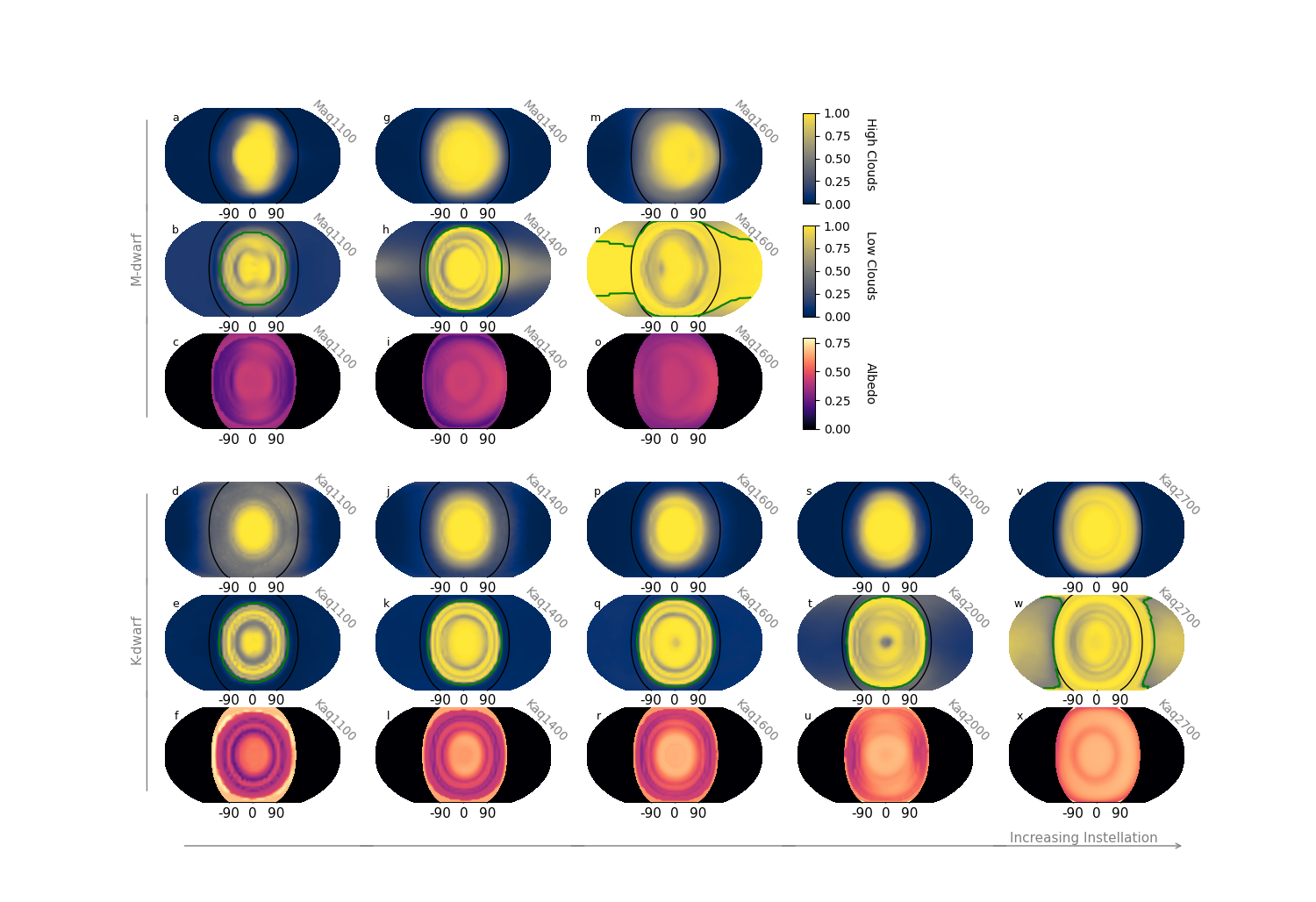}
\caption{\label{fig:Clouds} Cloud fraction and top-of-atmosphere albedo for M- and K-dwarf aquaplanets. From top to bottom, we show cloud fraction for high-level clouds (rows 1 and 4), cloud fraction for low-level clouds (rows 2 and 5), and top-of-atmosphere albedo (rows 3 and 6). Note that the peak in albedo near the terminator is due to sea ice. The green contours (rows 2 and 5) show sea ice edge, defined as the location where sea ice fraction surpasses 99\%.
}  
\end{figure}

It is well established that clouds play a critical role in M-dwarf planetary climate \citep{Yang2013, Yang2014lowordermodel}. In particular, the enhanced water vapor feedback in the upper atmosphere promotes nearly 100\% cloud fraction near the substellar point and high cloud fractions across the dayside (Fig.~\ref{fig:Clouds}). Water vapor clouds have a higher albedo on K-dwarf planets (e.g. Fig.~\ref{fig:Clouds} o and r). 
Comparing the Earth-like M- and K-dwarf planets (Maq1400 and Kaq1400), we note that the M-dwarf planet has a slightly less symmetric cloud coverage and additional equatorial nightside clouds. These small differences are expected given that our M-dwarf planet simulations are rotating faster than the corresponding K-dwarf planets, such that despite still being classified as slow rotators, they have a slightly stronger equatorial flow.

\begin{figure}[h]
\centering
\includegraphics[width=0.6\textwidth]{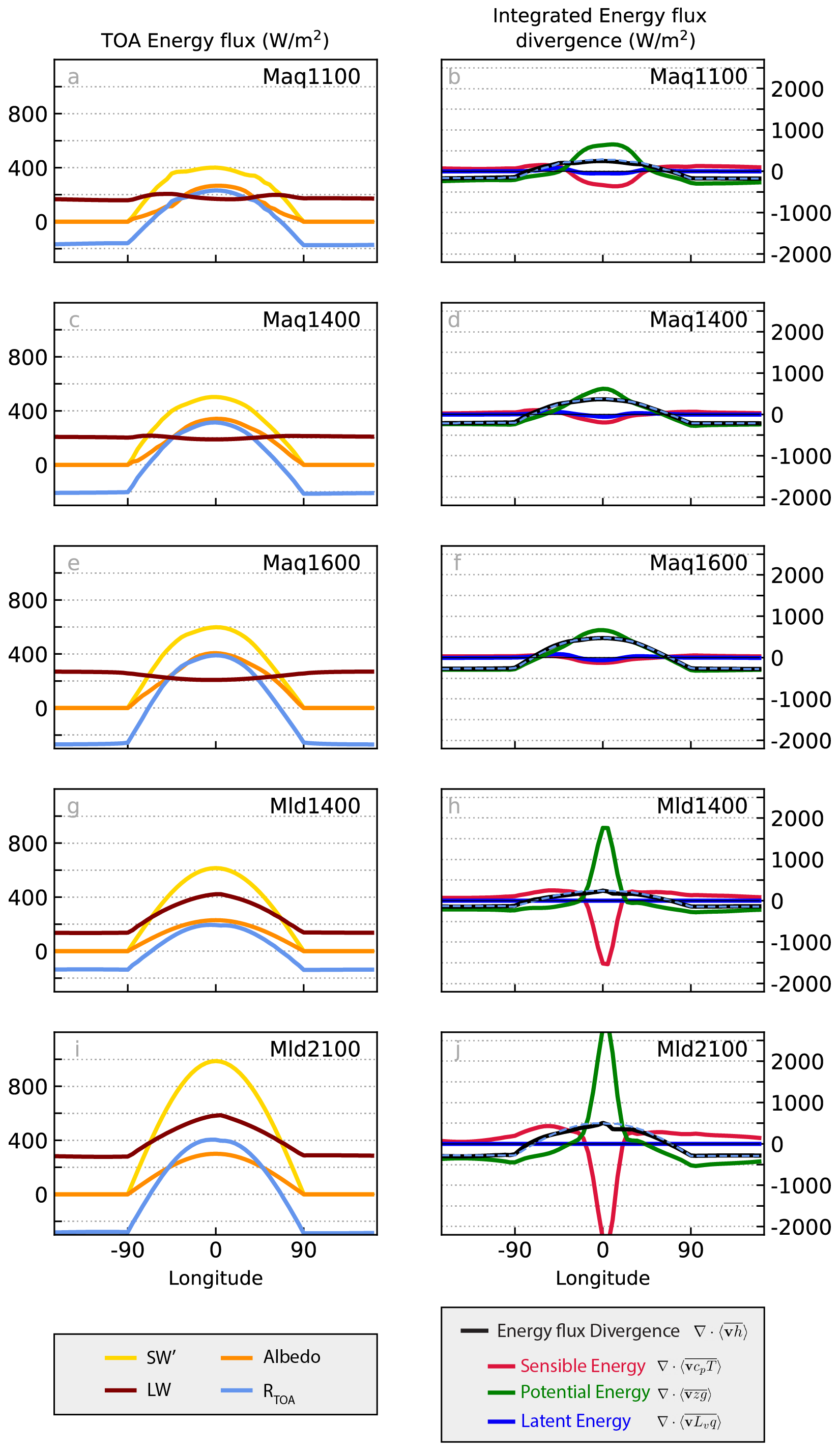}
\caption{\label{fig:NEI_M} 
Planetary energy budget for M-dwarf planets. The left column shows the radiative budget, including the TOA reflected stellar radiation (Albedo, orange), the incoming stellar radiation minus the reflected stellar radiation (SW', yellow), the net TOA thermal emissions (LW, maroon), and the net TOA radiation (R$_{TOA}$, light blue). The vertically integrated components of the energy flux divergence ($E = \nabla \cdot \langle \overline{\mathbf{v}h} \rangle$, black) are shown on the right, separated into sensible (red), latent (blue), and potential (green) energy components. The net TOA radiation is shown again (dashed blue line) and overlaps with the net energy flux divergence, indicating that $F_{\text{sfc}}$ is small. 
}  
\end{figure}

\begin{figure}[h]
\centering
\includegraphics[width=0.6\textwidth]{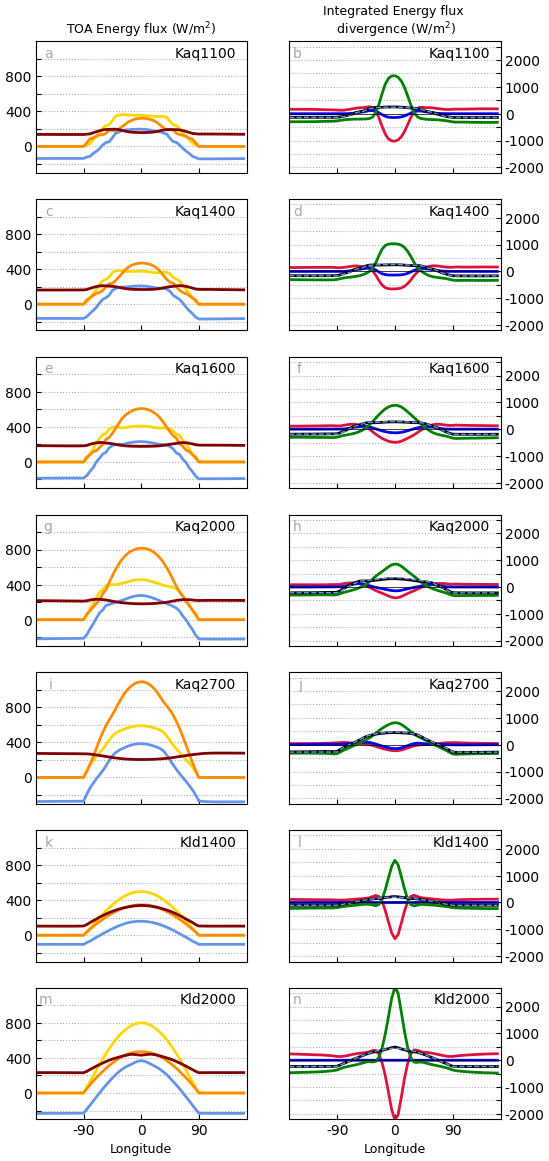}
\caption{\label{fig:NEI_K} 
Planetary energy budget for K-dwarf planets. The figure properties are the same as Fig.~\ref{fig:NEI_M}. 
}  
\end{figure}

All clouds have both a cooling and warming effect on climate, due to their contribution to the greenhouse effect and the planetary albedo \citep{CloudsAndClimate, Randall2012, Pierrehumbert2010}. However, higher clouds tend to have a greater warming effect because their cool cloud-tops result in a lower emission temperature. High clouds are concentrated in the substellar region, where the vertical branch of the overturning circulation is located and condensation is helping to drive the circulation (Fig.~\ref{fig:Clouds}, rows 1 and 4). These clouds remain  geographically confined to roughly the same area even as the orbital radius is reduced. Their impact on the planetary emissions can be noted in Figs.~\ref{fig:NEI_M} and \ref{fig:NEI_K}, where the aquaplanets' thermal emissions (maroon line) have a local minima in the substellar region. 

Meanwhile, low-level clouds have a net cooling effect due to their high albedo and their relatively high emission temperature leading to a smaller contribution to the greenhouse effect. The low level cloud coverage varies more dramatically as a function of orbital radius than the high-level clouds. Relative to the cooler planets (aq1100), we note that as orbital radius is decreased, low-level cloud coverage also expands from the substellar region towards the terminator. As planets approach the inner habitable zone edge, we note a slight increase in nightside low-level clouds, for both M- and K-dwarf aquaplanets. Though the nightside cloud fraction remains relatively low until the planets reach the MGL (Maq1600 and Kaq2700). Also, for both M- and K-dwarf planets, the MGL occurs at significantly higher instellation values than predicted in \citet{Kopparapu2013}, which did not account for the effects of clouds.

Overall we note that the cloud region, especially for low-level clouds, closely follows the sea ice edge. For an aquaplanet, this means that observations of water vapor clouds could potentially provide a rough indication of surface temperature patterns below. Though, of course, we are not yet accounting for other potential sources of clouds and hazes, as well as the impact of continents. 
With these caveats in mind, we also highlight that the terminator region remains relatively cloud free for simulations below the MGL. Given that clouds can hinder spectral measurements \citep[][]{Fauchez2019, Komacek2020, Suissa2020}, the terminator region may be easier to sample. 

\begin{figure}[h]
\centering
\includegraphics[width=0.6\textwidth]{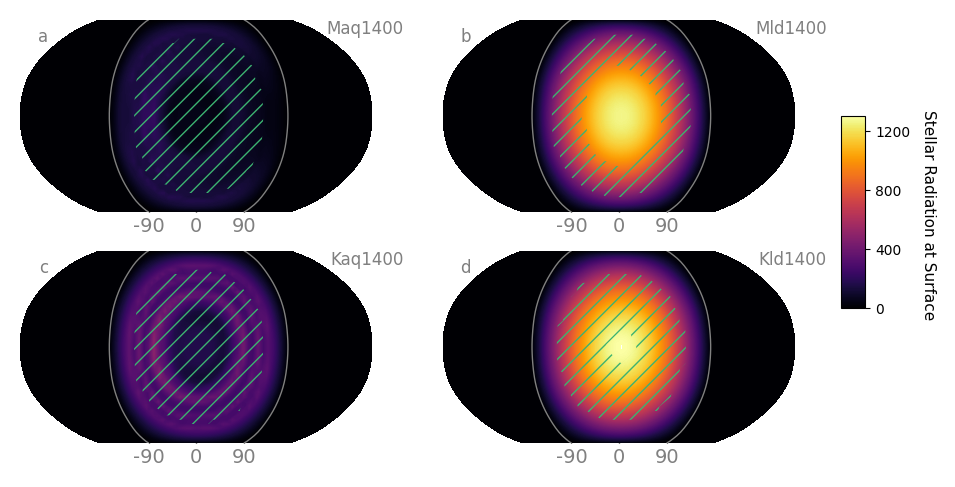}
\caption{\label{fig:sw_and_hab} Stellar radiation (W/m$^2$) reaching the surface of Earth-like M- (top) and K- (bottom) dwarf planets, including both aquaplanets (left) and land planets (right). Habitable surface areas, where surface temperatures are between 0 and 50\textdegree C, are marked with green hatches. }   
\end{figure}

Clouds could also directly impact habitability. For example, we note that high cloud fractions substantially reduce the amount of radiation reaching the planetary surface. This is especially significant for M-dwarf planets, where the atmospheric moisture not only results in increased albedo due to clouds, but also leads to the absorption of a significant portion of incoming shortwave radiation by the upper atmosphere \citep{Lobo2023}. For an Earth-like planet (Maq1400), less than 50 W/m$^2$ of stellar radiation reach the substellar planetary surface region (Fig.~\ref{fig:sw_and_hab}, a). This means that the planet's dayside surface is effectively dark. For K-dwarf planets, the intensity of this effect is reduced but still significant. A comparable planet (Kaq1400) receives 166 W/m$^2$ at the substellar point (Fig.~\ref{fig:sw_and_hab}, c). This is still relatively low compared to daytime values on Earth, but is more than triple the M-dwarf value and a greater fraction of that radiation would be in wavelengths that we know are viable for photosynthesis (visible spectrum). Therefore, it could be argued that the K-dwarf substellar region may be a more promising environment for complex life as we know it. Also, on both aquaplanets, surface instellation values are higher near the terminator than at the substellar region. For example, at 75\textdegree \space from the substellar point, their respective surface instellation values are 160 W/m$^2$ and 267 W/m$^2$, such that the terminator region may offer additional advantages for life relative to the substellar region.

\clearpage 

\clearpage

\subsection{Day-night Temperature Contrast}\label{sub:delT}

\begin{figure}[h]
\centering
\includegraphics[width=0.6\textwidth]{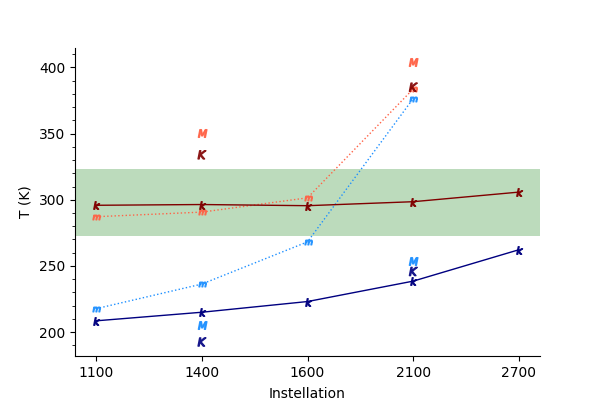}
\caption{\label{fig:MinMaxT} Minimum (blues) and maximum (reds) surface temperatures. Dotted line connects values from M-dwarf aquaplanet simulations, and solid line connects values from K-dwarf aquaplanet simulations. The capital M and K markers show land planet values. The shaded region highlights the temperature range between 0\textdegree C and 50\textdegree C where we might expect climate favorable to complex life forms. }   
\end{figure}

Comparing aquaplanet simulations, we note that K-dwarf planets overall have greater day-night temperature gradients than their M-dwarf counterparts. In both cases, these gradients decrease for planets closer to their host star. Prior to reaching the MGL, the decrease in gradient is almost entirely due to a rise in nightside temperatures. On K-dwarf aquaplanets, as the atmospheric water vapor increases, so does the cloud coverage, resulting in higher dayside albedos. For a large portion of the habitable zone, increasing dayside albedo is able to largely compensate for increasing stellar radiation. This can be noted in Fig.~\ref{fig:NEI_K}, by comparing the reflected stellar radiation (orange, A) and the remaining incoming stellar radiation (yellow, SW'), which quantifies the energy that will actually interacts with the planetary climate. The K-dwarf energy budgets (Figs. \ref{fig:NEI_M} and \ref{fig:NEI_K}) show that the amount of stellar radiation entering the climate system remains nearly constant across K-dwarf aquaplanets. The impact on dayside climate can be noted in Fig.~\ref{fig:MinMaxT}, which shows that K-dwarf dayside temperature maxima are nearly identical across simulations. 

\begin{figure}[h]
\centering
\includegraphics[width=1\textwidth]{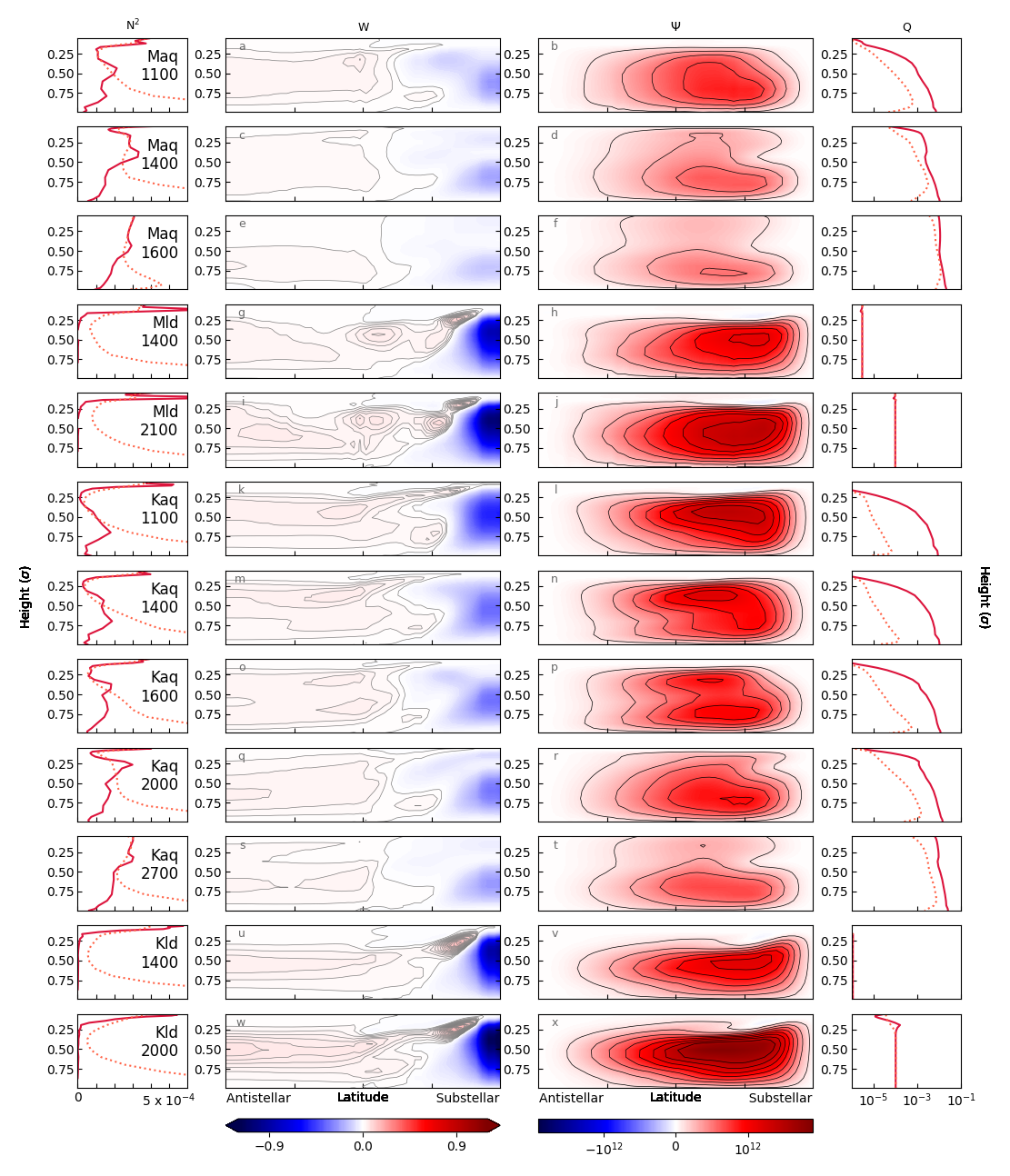}
\caption{\label{fig:strm} Vertical profile of M- and K-dwarf aquaplanets and land planets, with each row depicting a different simulation. The far left column shows vertical profile of the Brunt–Väisälä frequency squared (1/s$^2$) at the substellar (solid) and antistellar (dashed) points. Large values indicate regions of greater stability. Center left column shows vertical velocities (Pa s$^{-1}$). To facilitate visualization, grey contours were added to the region of descending air in the nightside. Center right shows stream functions ($\Psi'$, kg s$^{-1}$). Note the central two columns were calculated using the ``Tidally Locked" projection, taking a zonal average around the substellar point. The far right column shows vertical moisture profile (kg/kg) at the substellar (solid) and antistellar (dashed) point.}     
\end{figure}

Nightside temperatures are lower on K-dwarf aquaplanets than M-dwarf planets. For planets closer to their host star, the nightside becomes warmer partially due to an increase in energy flux convergence (Fig.~\ref{fig:NEI_K}). Note that the increase is very small and barely visible in the plots. The increase in nightside temperatures is also partially due to increased atmospheric moisture which amplifies the nightside greenhouse effect. Nightside vertical moisture profiles are shown in Fig.~\ref{fig:strm} (right column, dashed red lines). 

For M-dwarf planets, changes in albedo (Fig.~\ref{fig:Clouds}) and reflected light are smaller, such that there is a more significant increase in SW' as orbital radius is decreased (Fig.~\ref{fig:NEI_M}), which is balanced by increased day-to-night energy transport (Section \ref{sec:Overturning}). As we can note in Fig.~\ref{fig:MinMaxT} and Table \ref{tab:T_values}, dayside temperature changes are also small for M-dwarf aquaplanets until they reach the MGL, but the nightside temperature minima are rising roughly three times faster than on K-dwarf planets.

\section{Vertical Structure of the Overturning Circulation}\label{sec:Overturning}

The general structure of the overturning circulation is similar across these planets, with ascending air concentrated at the substellar point. The near surface flow converges in the substellar region, and the upper level flow is diverging from the substellar region and sinks as it flows towards the nightside and cools. The overturning cell in each simulation is plotted in Fig.~\ref{fig:strm}, which shows the stream functions calculated using the ``Tidally Locked" projection. 

The width of the cells are comparable across simulations, for M- and K-dwarf planets. However, the intensity of the circulation varies substantially. The M-dwarf aquaplanets have a notably weaker overturning circulation in terms of mass flux. This trend was previously identified in \citet{Lobo2023}, which showed that M-dwarf aquaplanets have a weaker overturning circulation than corresponding land planets, yet have a more efficient day-to-night energy transport. Here we find a similar trend applies for K-dwarf planets. K-dwarf land planets have an overturning circulation that is comparable to or stronger than their aquaplanet counterparts (e.g. Fig.~\ref{fig:strm}) and the resulting land planet atmospheric energy transport is weaker (Fig.~\ref{fig:NEI_K}, c and k) than their aquaplanet counterparts. 

The energy budget for a synchronously rotating planet can be written as: 
\begin{equation} \label{eq: energy_vint}
     R_{\text{TOA}} - F_{\text{sfc}} = \nabla \cdot \langle \overline{\mathbf{v}h} \rangle,
\end{equation}
where $R_{\text{TOA}}$ is top-of-atmosphere radiative fluxes, and $F_{\text{sfc}}$ the surface fluxes, which include radiative, sensible, and latent heat flux at the surface. The moist static energy term is defined such that $h = c_{p} T + L_v Q + g Z $, comprised of dry enthalpy $c_{p} T$, latent energy $L_v Q$, and potential energy $gZ$. $\langle \,\cdot \, \rangle$ denotes a vertical integral, and $\overline{(\, \cdot \,)}$ denotes a temporal long-term average. These components are shown on the right columns of Figs.~\ref{fig:NEI_M} and ~\ref{fig:NEI_K}, where we provide the meridionally averaged values centered about the substellar point.

Comparing Kaq1400 and Kld1400, we can confirm that the land planet has weaker nightside energy convergence and lower dayside peak divergence (Fig.~\ref{fig:NEI_K}, c and k). The same is true for Kaq2000 and Kld2000 (g and m). Though these trends appear small when plotted as part of the energy budget, they lead to roughly a doubling of the day-night temperature contrast. 

Comparing the energy budget for M- and K-dwarf aquaplanet simulations, we can also more quantitatively note the impact of clouds. Among our K-dwarf aquaplanet simulations, decreasing orbital radius has a smaller impact on the amount of energy entering the system (yellow lines, Fig.~\ref{fig:NEI_M} and \ref{fig:NEI_K}) due to increased dayside albedo, both in terms of magnitude at the substellar point and in terms of geographical area covered by clouds (note the widening of the orange peaks in Fig.~\ref{fig:NEI_K}). Therefore, even though Kaq1100 and Maq1100 have similar energy budget terms, the remaining aquaplanet simulations with reduced orbital radii respond with distinct mechanisms where M-dwarf aquaplanets have increased energy transport and K-dwarf aquaplanets have increased dayside albedo. This distinction emphasizes that the weaker cloud albedo feedback on M-dwarf aquaplanets is tied to the weaker day-night temperature contrasts.

For both M- and K-dwarf aquaplanets we note a weakening of the overturning circulation as the planet approaches the star. The trend can also be noted in a reduction of velocities in the vertical flow. The core of the cell shifts downward and the circulation becomes more concentrated in the lower layers of the atmosphere. As can also be noted in Fig.~\ref{fig:strm}, the warmer planets have a stronger atmospheric stratification, which is indicated by the higher positive values of the Brunt–Väisälä frequency (N). In Fig.~\ref{fig:strm}, we plot N$^2$ (left column) to allow for more clear visualization of upper atmospheric patterns. We note that this figure does not capture the full range of near-surface values, with antistellar values exceeding our upper limits, and substellar values in land planets dropping below zero. The increased stratification occurs as temperature decreases more slowly with height. As the upper layers of the atmosphere warm up, the saturation vapor pressure increases and we see a corresponding increase in moisture values in the mid levels (right column). Note that in the substellar region, the lower atmosphere is always nearly at saturation (not shown). 

\begin{figure}[h]
\centering
\includegraphics[width=0.80\textwidth]{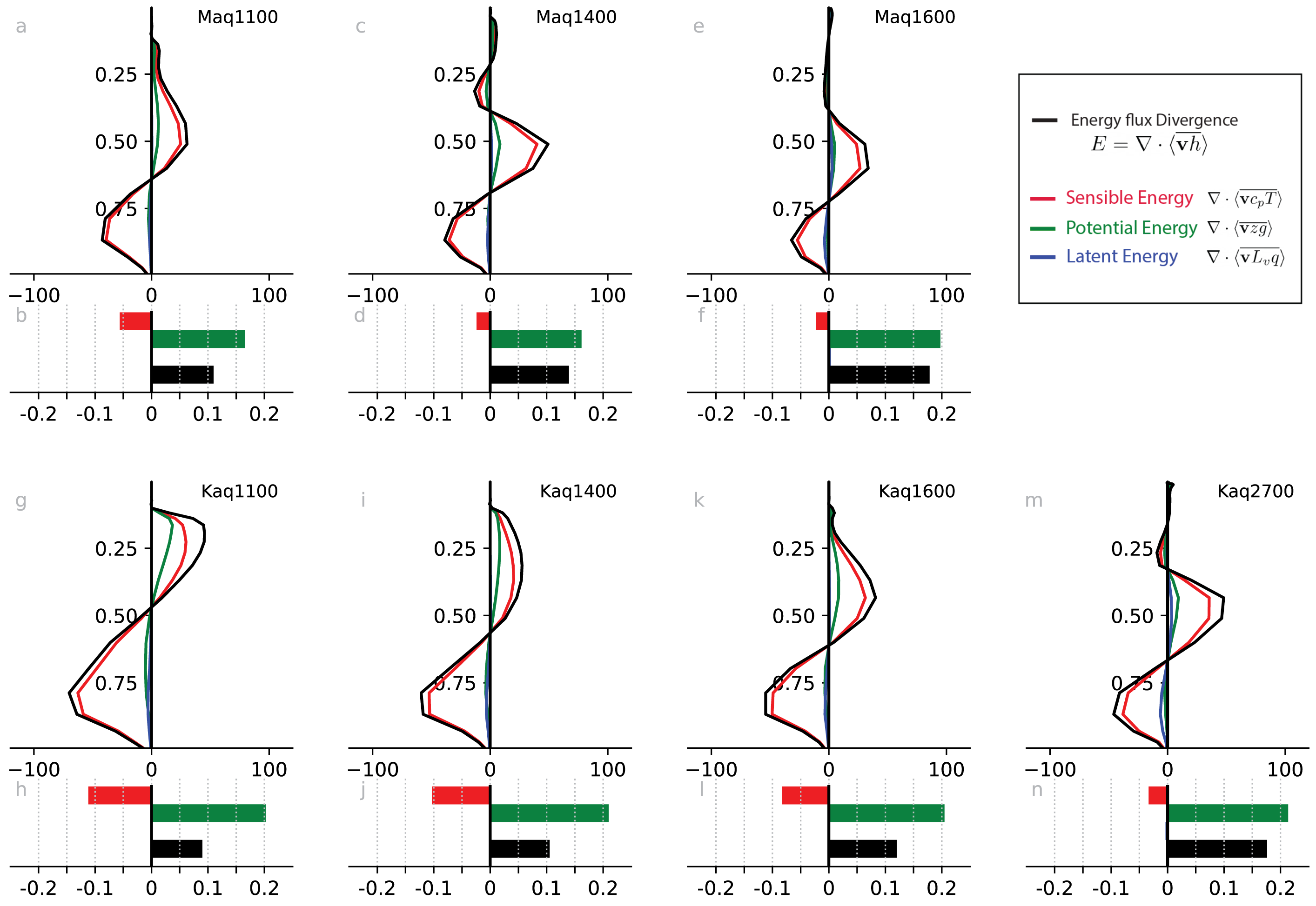}
\caption{\label{fig:NEIvertK} Energy budget for M- and K-dwarf aquaplanets. Line colors match Fig.~\ref{fig:NEI_K}. Top panels show vertical structure of the energy transport in the substellar region, plotting energy flux (x-axis, W/m$^2$) vs height (y-axis, $\sigma$). Values were averaged within a 15\textdegree \space radius of the substellar point. Lower panels show vertically integrated values averaged over the entire dayside, capturing the global day-to-night transport behavior. The x-axis shows vertically integrated energy flux, and each bar shows values for the energy flux components and total. When averaged over the entire dayside, latent energy values are too small to be visualized in this plot.}   
\end{figure}

\begin{figure}[h]
\centering
\includegraphics[width=0.750\textwidth]{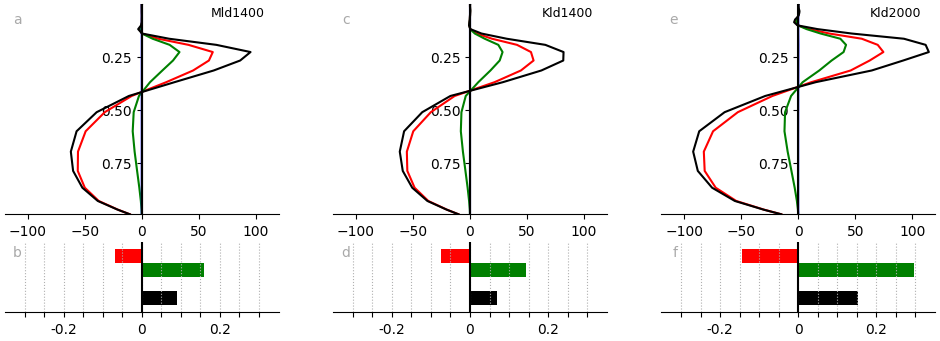}
\caption{\label{fig:NEIland} Energy budget for M- and K-dwarf land planets. The figure properties are the same as Fig.\ref{fig:NEIvertK} for land planets.  }   
\end{figure}

As a greater portion of the circulation becomes concentrated lower in the atmosphere, we note an overall reduction in the gravitational potential energy fluxes and sensible heat fluxes (Figs.~\ref{fig:NEI_M} and \ref{fig:NEI_K}, right). The decrease in sensible heat convergence is greater than the change in gravitational potential energy fluxes, such that the overall transport is still increasing as orbital radius decreases. This can be seen more clearly in Fig.~\ref{fig:NEIvertK}, by comparing values for Kaq1400 and Kaq2700. The decrease in sensible energy flux is not due to overall weaker sensible heat flow (Fig.~\ref{fig:NEIvertK}, i and m), but due to an increase in the sensible energy transported by the upper cell branch relative to the lower branch, such that it effectively cancels out most of the lower branch's transport, resulting in a reduced net sensible energy transport (panels j and n). For M-dwarf planets, this ``lowering" of the atmospheric circulation happens at lower instellation, with Maq1100 appearing to be in an intermediate regime with a circulation structure comparable to Kaq1600. Meanwhile, the circulation on cooler planets (Kaq1100 and Kaq1400) has a vertical structure that is qualitatively similar to land planets (Fig.~\ref{fig:NEIland}), though weaker in terms of magnitude.
Though they may appear subtle, as we shall discuss in the following section, these structural changes can have a significant impact on the planetary hydrological cycle. 

For land planets, the circulation vertical structure does not change substantially as we reduce the orbital radius (e.g. Fig.~\ref{fig:strm}, v and x). Unlike the aquaplanets, both the circulation (mass flux) and the net energy transport become stronger. There is a comparable strengthening of both the sensible and gravitational energy components, but the greatest increase occurs for the gravitational component in the upper branch of the overturning circulation, such that there is a net increase in energy transport (Fig.~\ref{fig:NEIland}, d and f).

These changes in the energy budget can have an important role in helping us understand planetary emission patterns. Almost all of our aquaplanet simulations, for M and K-dwarf planets, have comparable thermal emission at the substellar point. For aquaplanets that have not reached the MGL, thermal emission maxima occur near the terminator (Fig.~\ref{fig:NEI_M} a,c and \ref{fig:NEI_K}, a,c,e,g). Only the planets that have reached the MGL have nightside thermal emission maxima. This occurs on Maq1600 and Kaq2700. The Kaq2000 simulation is close to but not quite at the MGL and has a nearly flat emission profile. 

The difference between aquaplanet and land planet emissions are even greater. Unlike aquaplanets, land planet emissions peak in the substellar region and have nightside minima. The day-night emission gradient is also more than doubled in magnitude. Therefore, as observing capabilities for thermal emission phase curves advance, land planets may be especially convenient targets. Though in a more distant future, the terminator emission maxima could potentially be useful for linking emission patterns to the overall climate.

\section{Terminator Habitability on K-dwarf Planets} \label{TH}

We have defined ``terminator habitability" as a climate regime where a temperate habitable region exists near the terminator of a synchronously rotating planet, while dayside temperatures exceed habitable limits and nightside temperatures are below freezing. We also define habitability as ranging from 0\textdegree \space to 50\textdegree C. These limits are still subject to debate, but the trends we note for fractional habitability (Table \ref{tab:T_values}) are not strongly dependent on the exact limits. For M-dwarf planets, \citet{Lobo2023} showed that terminator habitability could occur for land planets but not aquaplanets. For K-dwarf planets, we find that terminator habitability is also not possible for aquaplanets. As can be noted in Fig.~\ref{fig:MinMaxT}, dayside temperatures on K-dwarf planets remain temperate and nearly constant even as the incoming stellar radiation is more than doubled. Even in Kaq2700, which has reached the MGL and has a nightside just on the cusp of melting, dayside temperatures remain mild. As with the M-dwarf aquaplanets, the day-night temperature gradients decrease as orbital radius decreases, such that the planetary climate becomes homogeneous before temperatures rise above habitability limits. Therefore, it is not possible for these Earth-like aquaplanets to cross into a terminator habitability regime. 

As with M-dwarf land planets, terminator habitability does occur on K-dwarf land planets. Kld2000 gives a clear example of terminator habitability, with dayside temperatures reaching a scorching 112\textdegree C, and a smaller habitable band occupying roughly 38\% of the dayside surface area (Fig.~\ref{fig:surfaceTemperature}l). The Kld1400 simulation is roughly at the transition between what we consider a habitable ``eyeball" climate and a "terminator habitability" regime. On this planet the habitable region covers the majority of the dayside surface, but does not include a small region at the substellar point where temperatures reach up to 61\textdegree C (Fig.~\ref{fig:sw_and_hab}d). Both M-dwarf and K-dwarf land planets in the inner habitable zone could be in terminator habitability regimes. Based on our sandy land planet simulations, in the outer habitable zone we might expect fewer K-dwarf land planets in a terminator habitability regime due to generally cooler dayside temperatures. But this result is dependent on planetary surface properties and a land planet with a darker surface (e.g. basaltic rock) could sustain terminator habitability conditions at greater orbital radii.

\begin{deluxetable}{cccllc}[h]
\tablenum{3}
\tablecaption{Planetary surface temperatures and fractional habitability (FH).  \label{tab:T_values}}
\tablewidth{0pt}
\tablehead{
\colhead{ }  & \colhead{Mean T} & \colhead{T$_{\text{max}}$ }
& \colhead{T$_{\text{min}}$} & \colhead{$\Delta$T} & \colhead{FH 0 - 50\degree C} 
}
\startdata
\multicolumn{6}{c}{M-dwarf planets} \\
Maq1100            &   245K &   288K &  216K &   72K  &   24\% \\
Maq1400    &   259K &   291K &   236K &   54K  &   32\%    \\
Maq1600    &   283K &   301K &   268K &   33K  &   79\%    \\
Mld1400  &   247K &   350K &   205K &   145K &   24\%    \\
Mld2100  &   295K &   404K &   254K &   150K &   16\%    \\
\multicolumn{6}{c}{K-dwarf planets} \\
Kaq1100 & 238K & 296K & 209K & 87K  & 26\%  \\
Kaq1400 & 246K & 296K & 215K & 81K  & 31\%  \\
Kaq1600 & 252K & 296K & 223K & 72K  & 34\%  \\
Kaq2000 & 262K & 299K & 238K & 60K  & 39\%  \\
Kaq2700 & 280K & 306K & 259K & 47K  & 52\%   \\
Kld1400 & 234K & 334K & 193K & 141K & 29\%  \\
Kld2000 & 282K & 385K & 245K & 139K & 19\%  \\
\enddata
\end{deluxetable}

\section{The Water Cycle and Long Term Water Availability on K-dwarf Planets}\label{WaterCycle}

Thus far we have defined habitability as a function of surface temperature. However, liquid water is a key ingredient for life as we know it and we must take extra care to evaluate the long term water availability on the planets we simulate. 
Aquaplanet simulations provide the atmosphere with an unlimited water source, such that even if a region is continuously losing water it never becomes dry. In order to evaluate the long term water availability, we cannot simply look at the simulated climate, but must also look at the trends in the moisture budget to determine how the planetary climate would evolve over time. For simplicity, we will discuss water budget trends in this section while neglecting Maq1600 and Kaq2700, which are past the MGL and would be at risk for atmospheric water vapor escape. 

We find that the K-dwarf aquaplanets simulations have higher precipitation rates than their M-dwarf aquaplanet counterparts, with maxima as high as 28.6 mm/day (Kaq1100). But, while we might have expected increased precipitation rates as orbital radius is reduced, we notice only slight variations in precipitation rates and no strong trends. We expect this may be due to changes in the vertical stratification (see section \ref{sec:Overturning}), such that the saturation vapor pressure is decreasing more slowly with height. Note that saturation vapor pressure has an exponential relationship with temperature, such that a small change in the vertical temperature structure leads to a large change in the atmosphere's water holding capacity. 

\begin{figure}[h]
\centering
\includegraphics[width=0.6\textwidth]{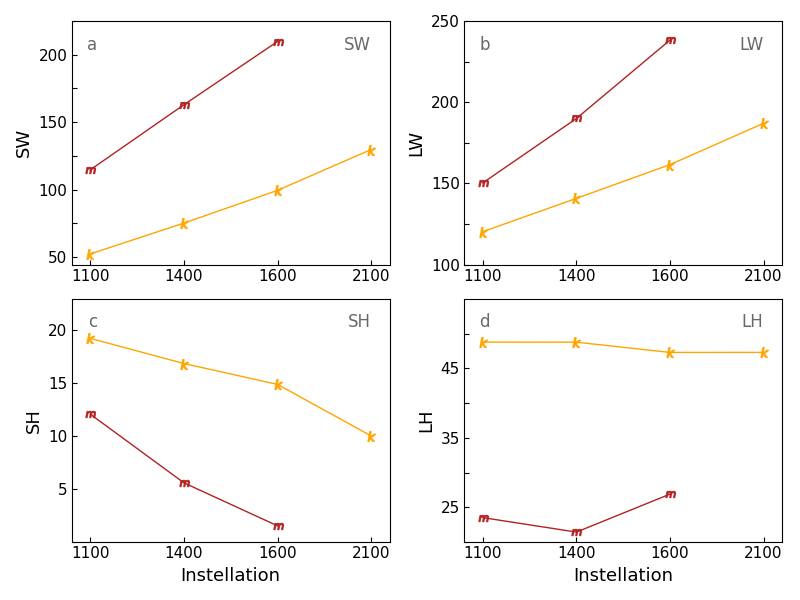}
\caption{\label{fig:precip_energy_budget} Comparison of global-mean energy budget components for M- and K-dwarf aquaplanets (red and yellow respectively). Panels show (a) atmospheric shortwave absorption, (b) atmospheric net longwave emission, (c) surface sensible heat flux, (d) surface latent heat flux, with all values in W/m$^2$.}    
\end{figure}

The behavior of dayside precipitation rates can be further understood through energetic constraints
\citep[e.g.][]{Ogorman2012, Allen2002, Xiong2022}. For global mean values, a budget can be written as:

\begin{equation}
    LH = LW - SW - SH
\end{equation}
where the surface latent heat flux (LH) is equal to atmospheric net longwave emission (LW), atmospheric shortwave absorption (SW), and surface sensisble heat flux (SH). See \citet{Xiong2022} for details. Given that these are global averages, the surface latent heat flux is equivalent to surface precipitation values multiplied by water's specific heat of vaporization and density. Comparing the budget for M- and K-dwarf planets (Fig.~\ref{fig:precip_energy_budget}), we find that M-dwarf planets have overall higher values of LW and lower values of SH, which would both favor increased precipitation. However, the greatest difference between the M- and K-dwarf simulations occurs for values of SW, where M-dwarfs have substantially greater values of absorption. Thus, we can attribute the overall reduced M-dwarf precipitation rates to the increased shortwave absorption, which is enhanced for M-dwarf planets due to the stellar emissions peaking in the near-IR.  

\begin{deluxetable}{ccc}[h]
\tablenum{4}
\tablecaption{Snow rates and Cold-trapping timescale.  \label{tab:snowrates}}
\tablewidth{0pt}
\tablehead{
\colhead{ }  & \colhead{Nightside Snow Rate (10$^{15}$ kg/year)} & \colhead{ Cold-trapping timescale (kyr) }
}
\startdata
Maq1100  & 4.9 & 271  \\
Maq1400  & 12.7 & 105    \\
Kaq1100  & 0.12 & 10,700 \\
Kaq1400  & 0.92 & 1449 \\
Kaq1600  & 3.02 & 443 \\
Kaq2000  & 7.19 & 186 \\
\enddata
\end{deluxetable}

\begin{figure}
    \centering
    \includegraphics[width=1\textwidth]{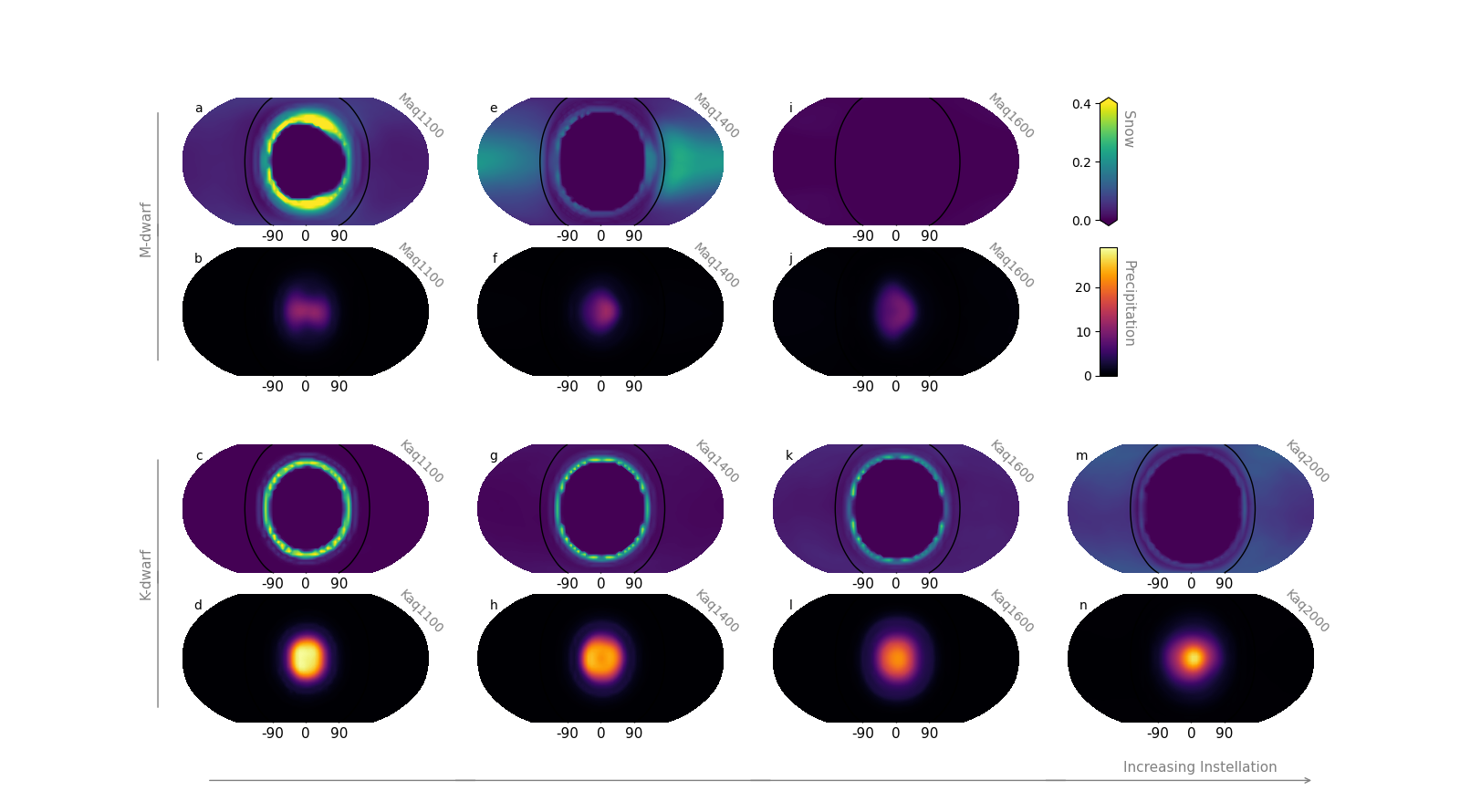}
    \caption{M-dwarf and K-dwarf planetary snow rates (top) and precipitation rates (bottom) in mm/day. Precipitation rates include both liquid and solid precipitation.  }
    \label{fig:SnowAndRain}
\end{figure}

Given that the air near the substellar surface is nearly saturated in all aquaplanet simulations, and the dayside holding capacity is increased without a comparable increase in precipitation, it follows that there should be an increase in day-to-night moisture transport for planets with smaller orbital radii. We confirm that moisture transport indeed increases for planets closer to their host star. With increasing day-to-night moisture transport, we observe a corresponding increase in nightside snow rates for planets with reduced orbital radii (Fig.~\ref{fig:SnowAndRain}).  The changes in snow rate are also quantified in Table \ref{tab:snowrates}, which provides an area weighted sum of the nightside snow rate for each simulation. Many of these planets could be vulnearble to nightside cold-trapping, such that over time a large portion of their water reservoirs could become trapped in nightside ice deposits, making the dayside climate water-limited or even arid. In Table \ref{tab:snowrates}, we provide a cold-trapping timescale that quantifies the time it would take for the atmosphere to transport a volume of water equivalent to Earth's oceans from the dayside to the nightside of the planet. This is an underestimation given that we are neglecting ocean heat transport, as well as return mechanisms such as glacier flow, but it provides us with a sense of scale for the intensity of the day-to-night water transport. 

Nightside snow rates are greatest for the planets near the inner edge of the habitable zone. Meanwhile, the trapping timescale for planets further out in the habitable zone, such as Kaq1100, are two orders of magnitude greater. This implies that there could be an outer limit to nightside cold-trapping. In the case of Kaq1100, the timescale of approximately 11 million years is relatively short and could still result in a ``drying" of the dayside climate. But we have reason to expect that many planets with comparable or lower snowfall rates might not be subject to intense nightside cold-trapping. Note that on M-dwarf aquaplanets, snow rates also decrease with distance from the host star but are higher than their K-dwarf counterparts.

It is difficult to pinpoint an exact outer limit for nightside cold-trapping because the mechanisms that would return water from the nightside to the dayside are poorly constrained. For example, ice flow could play an important role, but would depend on topography, glacier dimensions, and many other planet-specific properties. Also, melt rates would be strongly dependent on geothermal heating. As argued in \citet{Yang2014watertrap}, higher geothermal heating rates on super-earths could help prevent nightside cold-trapping. However, it has also been proposed that planets larger than 2.5 Earth masses might not be able to sustain a strong dynamo \citep{Gaidos2010}, and the absence of a magnetic field would make such planets a less appealing target in the search for life. For now, we will assume Earth-like conditions and use Antarctica rates as a reference. 

Estimates of the geothermal heat flux at the base of the west Antarctica ice sheet imply basal melt rates as high as 0.049 mm/day \citep{Fisher2015}. Basal melt would not necessarily translate into a night-to-day flow, especially if glaciers only covered a portion of the planet's nightside. The west Antarctica ice sheet is also associated with high rates of melting even for Earth. But, for the sake of a simple comparison, we can treat it as an upper limit for the expected return flow and compare it to nightside snow rates. Such a high melt rate would exceed the average nightside snowfall for Maq1100, Kaq1100 and even Kaq1400. In particular, Kaq1100 has an average snowrate that is an order of magnitude smaller. With this in mind, for a planet similar to Earth, we would consider the risk of nightside cold-trapping at or beyond this orbit to be low.

The ocean circulation could also help reduce nightside cold-trapping by decreasing the fraction of the nightside ocean that is covered in sea ice, allowing for a direct mechanism to return water to the dayside. Of course, this assumes the presence of global-scale ocean basins. For example, planets with a ``lobster-like" configuration \citep{Hu2014} could accumulate snow more slowly, though the magnitude of the ocean circulation's effect would depend on ocean properties including salinity and basin depth \citep{Olson2020}. Also, \citet{Yang2019oceandyn} showed that ocean energy transport can play a significant role in the day-to-night energy transport for M-dwarf planets in the outer habitable zone, but becomes negligible for planets in the inner habitable zone, likely due to weak wind stress and weak shortwave radiative forcing at the surface. Additional analysis would be need to explore the implications, but such a trend in ocean energy transport could serve to further reduce cold-trapping risks in the outer habitable zone while leaving the inner region vulnerable. With this in mind, we are likely underestimating the cold-trapping timescales and the outer limit of the cold-trapping region may be closer to the host star. 

Comparing M- and K-dwarf simulations, we also note that K-dwarf planets and Maq1100 have a high dayside snow rates (Fig.~\ref{fig:SnowAndRain}), which forms a ring-like shape near the terminator. The dayside snowfall occurs just inward of the sea-ice edge and is not directly related to the previously discussed day-to-night moisture transport. Instead, it is a result of low level atmospheric turbulence which leads to condensation in the lower atmospheric layers (below 0.8 $\sigma$). The condensation is recorded as snowfall, but does not lead to surface ice accumulation because it is falling on the open ocean. Since this water would remain an active part of the water cycle, this feature's impact is limited for our aquaplanets. We also do not observe these features in our land planet simulations which are arid and have minimal precipitation. But, it is possible that on more water-rich land planets this feature could provide a mechanism for retaining water and ice reservoirs on the dayside and closer to the terminator. 

\section{Discussion} \label{Discussion}

The climate trends we have explored are qualitatively similar for M- and K-dwarf habitable zones. However, the quantitative differences in these trends imply significantly different distribution of climate regimes across their habitable zones. For example, terminator habitability is possible on both M- and K-dwarf land planets. However, the regions where terminator habitability is likely to occur are reduced for K-dwarf planets, and more narrowly confined to the inner portion of the habitable zone. For aquaplanets, high nightside snow rates could result in rapid nightside cold-trapping for planets in the inner habitable zone of both M- and K-dwarf stars. But nightside snow rates are overall lower for K-dwarf planets, such that a smaller portion of their habitable zone would be at risk of cold-trapping. Based on this result, the K-dwarf inner habitable zone would be more likely to have planets in an ``eyeball" climate regime with a moist dayside climate. 

Previous work exploring M-dwarf planetary climate showed that, even if M-dwarf planets formed with abundant water, atmospheric escape and nightside cold-trapping could make it challenging for synchronous planets to retain their dayside water reservoirs. Comparing Maq1400, which has a short nightside cold trapping timescale, and Maq1600 which surpasses the MGL and is at risk of water vapor escape, we note that there is little or no gap in orbital radii parameter space where an Earth-like planet would not be subject to at least one of these effects. However, it is possible that a broader gap exists for planets with different bulk atmospheric compositions or greater geothermal heating. We also have not yet explored planets with intermediate levels of water availability. On a water-rich land planet, it could also be possible that previously neglected effects such as topography and surface water distribution could slow the day-to-night moisture transport, leading to a potentially broader gap. 

While an Earth-like atmosphere is able to absorb a large fraction of incoming M-dwarf stellar radiation, which can serve as a significant heat source in the upper atmosphere, the absorption of incoming K-dwarf radiation is much weaker. Therefore, we expected the inward shift of the habitable zone edge (defined as a function of instellation) to allow for a greater gap between orbital radii subject to nightside cold-trapping and those reaching the MGL. However, we find that for an Earth-like aquaplanet, this is not the case. Instead, nightside cold-trapping timescales decrease as orbital radius is decreased, such that a gap would only be possible if the nightside were to melt before reaching the MGL, and that does not occur for our Earth-like aquaplanets. Even Kaq2700, which as reached the MGL, still has a frozen nightide. Therefore, for an Earth-like planet, dayside water retention in the inner habitable zone could be less likely, and would be dependent on the presence of additional mechanisms, such as strong ocean energy transport. 

For our Earth-like synchronously rotating planets, these result suggests that the mid and outer habitable zone could be a more promising region to search for potentially habitable and water abundant planets with moist dayside climates. Meanwhile, water-limited land planets could be common across a broader portion of the habitable zone. Also, planets that were initially ocean-covered worlds subject to strong nightside cold-trapping could be considered water-limited land planets once most of their water reservoir is in the form of ice. If located in the outer habitable zone, these water-limited planets would tend to be in a habitable ``eyeball" climate regime. For those near the inner edge of the habitable zone, we'd expect many to be in a ``terminator habitability" climate regime. 

A significant caveat to this assessment is that planetary properties, including atmospheric pressure and composition could significantly impact the habitable zone limits, and the regions where we'd expect to find each climate regime. The inclusion of an active carbonate-silicate cycle could, for example, play a large role in determining whether a land planet will be in a ``terminator" or ``eyeball" habitability regime. Also, we have kept atmospheric surface pressure constant and Earth-like in order to focus on the role of the stellar SED. Previous work has showed that background gas pressure has a relatively small impact on the inner edge of the habitable zone \citep{Zhang2020}, and found that surface temperatures remained almost unchanged when the surface atmospheric pressure is halved \citep{Lobo2023}. However, the impact on the moisture budget is not necessarily negligible. For example, the nightside snow rates for a planet with half the atmospheric mass of Maq1400 is nearly doubled ($22 \times 10^{15}$ kg/yr). Therefore additional work is needed to quantify the impact of planetary properties on the water budget and climate regimes.

Regardless of which climate regime a planet is in, for these slow rotators, the terminator region has the lowest dayside cloud fraction. Recent studies have shown that clouds, much like hazes, could pose a challenge for spectral observations \citep{Fauchez2019, Komacek2020, Suissa2020}. Substantial cloud coverage can lead to a reduction of the relative transit depths, making it harder to characterize the atmospheric composition and potentially identify biosignatures. As argued in \citet{Lobo2023}, the terminator could be an especially promising region on water-limited planets in terms of water availability and habitability. In this work we find that remains true for both M- and K-dwarf synchronously rotating planets. Even for water-rich planets in a habitable ``eyeball" regime, the terminator zone (including the dayside ``mid-latitudes") could be especially advantageous for photo-dependent life because a much greater fraction of stellar radiation reaches the planetary surface. Also, the low cloud fraction at the terminator region is likely to facilitate observations, and in some cases, could facilitate the detection of shorter-lived and regionally concentrated biosignatures. With near-future observations in mind, we expect that planets with habitable regions near the terminator may be especially promising for detecting life, even if the planet is in a habitable ``eyeball" regime.

However, before finding a potentially ``inhabited" planet we are more likely to find ``habitable" planets. In which case, our primary focus would be on climate characterization and identifying planets with surface liquid water. With this in mind, the outer habitable zone is especially appealing. With a reduced cold-trapping timescale, these planets would be more likely to have dayside surface water, while also having reduced cloud coverage such that we could sample a greater portion of the dayside atmosphere. They would be less likely to have lost their atmospheres during the early stages of stellar evolution, as we believe may have happened for the inner TRAPPIST-1 planets \citep{Greene2023, Lincowski2023}. Therefore, as we plan for the next generation of space based telescopes \citep{luvoirfinalreport, habexfinalreport, decadalsurvey}, it would be of interest to include a greater sample of M- and especially K-dwarf planets in the outer habitable zone.


\section{Conclusions}\label{sec:Conclusions}

For synchronously rotating M- and K-dwarf aquaplanets, we find that the inner habitable zone is not only at greater risk of water vapor escape, but also has a higher risk of nightside cold-trapping due to enhanced day-to-night moisture transport. These results suggest that, especially for K-dwarf systems which have reduced nightside snow rates, the mid and outer habitable zones may offer better prospects for detecting water-rich worlds. Meanwhile the inner habitable zone would tend to have a higher concentration of water-limited planets, which could be particularly common for M-dwarf systems given the challenges discussed for obtaining and retaining water on M-dwarf planets. 

We show that water-limited planets can be in a terminator habitability regime, which is especially likely to occur in the inner habitable zone where dayside temperatures tend to exceed habitable limits. Though our simulated land planets are arid and arguably less Earth-like than their aquaplanet counterparts, the habitable regions on their surfaces are more Earth-like in terms of their surface instellation values. The reduced dayside cloud coverage and large day-night thermal emission gradients could also offer substantial observational advantages, making such water-limited planets especially promising near-future targets. 

\section{Acknowledgments}

This material is based upon work supported by the National Science Foundation under Award 1753373, the NASA Exoplanet Research program Award 80NSSC23K1069, and by a Clare Boothe Luce Professorship supported by the Henry Luce Foundation. This research was also performed as part of the NASA's Virtual Planetary Laboratory, supported by the National Aeronautics and Space Administration through the NASA Astrobiology Institute under solicitation NNH12ZDA002C and Cooperative Agreement Number NNA13AA93A, and by the NASA Astrobiology Program under grant 80NSSC18K0829 as part of the Nexus for Exoplanet System Science (NExSS) research coordination network.
We would like to acknowledge high-performance computing support from Discover, provided by the NASA Center for Climate Simulations (NCCS).

\bibliography{bibs}{}
\bibliographystyle{aasjournal}



\end{document}